# Origin of the electrocatalytic activity in carbon nanotube fiber counter-electrodes for solar-energy conversion

*Alba Martínez-Muíño[ab], Moumita Rana[a], Juan J. Vilatela[a]\* and Rubén D. Costa[ac]\**

Carbon nanotubes are a versatile platform to develop sustainable and stable electrodes for energy-related applications. However, their electrocatalytic activity is still poorly understood. This work deciphers the origin of the catalytic activity of counter-electrodes (CEs)/current collectors made of self-standing carbon nanotubes fibers (CNTfs) using $Co^{+2}/Co^{+3}$ redox couple electrolytes. This is based on comprehensive electrochemical and spectroscopic characterizations of fresh and used electrodes applied to symmetric electrochemical cells using platinum-based CEs as a reference. As the most relevant findings, two straight relationships were established: i) the limiting current and stability increase rapidly with surface concentration of oxygen-containing functional groups, and ii) the catalytic potential is inversely related to the amount of residual metallic Fe catalyst nanoparticles interspersed in the CNTf network. Finally, the fine tune of the metallic nanoparticle content and the degree of functionalization enabled fabrication of efficient and stable dye-sensitized solar cells with cobalt electrolytes and CNTf-CE outperforming those with reference Pt-CEs.

## Introduction

Carbon nanotubes (CNT) are an attractive material for sustainable and highly efficient electrodes in electrocatalytic energy-related applications.[1,2,3] However, there are still uncertainties regarding the origin of their catalytic activity with respect to surface functionalization, type of CNTs/entanglement, and catalyst impurities. For instance, CNTs are generally synthetized using metal catalyst, such as Fe, Cu, or Ni.[4–6] The residual catalyst can reach >30% wt.,[7] most often as metallic nanoparticles, but also as carbides and/or oxides.[7,8] They are either encapsulated in the CNT network or as interspersed nanoparticles capped by a graphitic shell. Up to date, the presence of metallic impurities has scarcely been recognized as critical for electrocatalytic processes.[9] In contrast, there has been much more research on the electrocatalytic behavior of CNTs in terms of type of CNTs,[3,10] functionalization degree and/or defects,[11] and CNT alignment.[12,13]

Despite huge efforts, the role of defects and impurities in CNT-based electrodes for solar energy conversion processes is still elusive.[1,2] This is especially critical for dye-sensitized solar cells (DSSCs), in which the electrochemical regeneration of the dye and the electrolyte are key processes.[14–15]

Up to date, the DSSC field has reached a mature state, achieving energy conversion efficiencies ($\eta$) superior to 14% (outdoor)[14] and 28.9% (indoor).[15] There is still extensive research focused on: i) increasing device efficiency and stability,[14,15] ii) replacing unsustainable materials, such as Pt at the counter-electrode (CE) and iodine electrolytes,[16–18] and iii) enabling augmented mechanical properties.[19,20] In this context, CNT electrodes are becoming key to tackle these challenges towards highly efficient DSSCs as they combine i) ultrafast exciton/charge transfer (1–10ps),[21–25] ii) extraordinary high mobility ($10^5 cm^2/Vs$),[26] iii) high electrochemical stability,[27–29] iv) mechanical features when assembled as macroscopic CNT fibres or fabrics (CNTf),[30,31] and v) simple, rapid, and inexpensive processing as large-area structures.

CNTs, and specially MWCNTs, substantially lower the charge-transfer resistance of the DSSC, along with a good electrode-electrolyte interface and a high specific surface area, which is mainly reflected in the FF and DSSCs efficiencies.[32,33] Indeed, wire-like DSSCs with CNTf-CEs and $TiO_2$ photoanode have reached $\eta$>10% as wearable devices.[31] In contrast, planar architectures had commonly achieved low efficiencies with a $I^-/I_3^-$ redox electrolyte,[34,35] until we have recently demonstrated that highly graphitic and crystalline CNTfs used as current collector/CE realizes $\eta$ of *ca*. 9%.[36] Improving the CEs' performances is still a challenge, and composites based on carbon materials (such as CNTs, carbon nanofibers-CNF, graphene, etc.) and transition metal compounds (TMCs) are actively researched candidate materials. These composites generally improve the electrochemical performance of the TMC and the electrocatalytic activity towards triiodide reaction, such as in the case of CNF-Pt composites.[37–39] On the other hand, they tend to present a lower charge-transfer resistance (compared to Pt) and high surface area,[40–44] derived from the carbon material.

Despite their promising performance as CE,[34–36,45–48] the origin of the catalytic behavior of CNTf-electrodes (or their composites) is also poorly understood in DSSCs. Current gaps in our understanding of the origin of the catalytic activity have, in addition, restricted their use to iodine-based electrolytes, when in fact carbon derivatives are known to be highly efficient catalysts for many other redox couples and hole transport materials (HTMs).[49–51]

In this context, this work sets out to decipher the origin of the catalytic behavior of CNTf in terms of surface functionalization and metallic impurities using emerging electrolytes ($Co^{+2}/Co^{+3}$ redox)[52] for highly efficient planar DSSCs. This rationale is supported by a comprehensive spectroscopic and electrochemical study of fresh and used CNTf-electrodes applied to both symmetric cells and fully operational DSSCs. As the most relevant findings, two straight relationships are established: i) the limiting current and stability increase dramatically with the degree of oxidative CNTf functionalization, and ii) the catalytic potential is inversely related with the amount of residual metallic Fe catalyst interspersed in the CNTf network. Hence, metallic impurities are the catalytic centers, while surface functionalization rules the limiting current. Based on these findings, we further optimized the CNTf-electrodes to achieve highly efficient and stable DSSCs, outperforming those with reference Pt-CEs.

[a] *IMDEA Materials Institute, c/ Eric Kandel 2, Getafe, 28906, Madrid, Spain E-mail: juanjose.vilatela@imdea.org, ruben.costa@imdea.org*

[b] *Universidad Autónoma de Madrid, Departamento de Física Aplicada, Calle Francisco Tomás y Valiente, 7, 28049, Madrid, Spain*

[c] *Technical University of Munich, Chair of Biogenic Functional Materials Schulgasse, 22, 94315, Straubing, Germany. E-mail: ruben.costa@tum.de*

Overall, this work clearly demonstrates that CNTf-CEs featuring metallic Fe impurities lead to highly performing multifunctional electrodes – *i.e.,* electrocatalytic electrolyte regeneration, and current collector – beyond using traditional redox couples in DSSCs. This paves the way towards future works focused on copper(I) complexes electrolytes [15,53], thiolate/disulphide [54–56] for tandem solar cells, and polysulfides for quantum dot sensitized solar cells.[57,58]

## Experimental

### Materials

Thiophene (extra purity ≥99%) and ferrocene (purity = 98%) were obtained from Acros Organics and 2-butanol (purity >99%) from Sigma Aldrich. Ferrocene was purified by a sublimation/recrystallization process. The D35 organic dye (DN-F04) and the [Co(bpy)$_3$][TFSI]$_2$/[Co(bpy)$_3$][TFSI]$_3$ redox couple used for the preparation of the cobalt-based electrolyte were purchased from Dyenamo AB (Sweden). 1-butyl-3-methylimidazolium iodide, guanidine thyoicianate, 4- tert-butylpyridine, iodine, acetonitrile (ACN) and valeronitrile were purchased from Sigma Aldrich. Transparent conductive glass substrates were obtained from XOP Glass (FTO, 15 Ω per square). TiO$_2$ 18NR-AO Active Opaque Titania Paste and 18NR-T Transparent Titania Paste was purchased from Dyesol UK. TiCl$_4$ was bought from Fisher Scientific (titanium (IV) chloride solution, 0.09 M in 20% HCl). CNTf with effectively 0% metal content was purchased from DexMat.

### CNTf synthesis

The CNTf were synthesized by the direct spinning floating catalyst chemical vapor deposition (CVD) method[59] using butanol as carbon source, ferrocene as Fe catalyst source and thiophene as promoter, at a concentration (0.8 wt.% of ferrocene, 1.5 wt.% of thiophene and 97.7 wt.% of butanol) adjusted so as to produce long, highly graphitic CNTs of few layers.[60] The reaction was carried out in hydrogen atmosphere at 1250 °C in a vertical furnace, using precursor feed rates 5 mL/h and a fiber winding rate of 7-9 m/min. Electrodes were produced by directly winding multiple CNT filaments on a paper substrate so as to form a unidirectional non-woven fabric. The fabric is a highly aggregated porous material with high mechanical robustness. This material was transferred to FTO-glass using a simple press-transfer technique to fabricate electrodes with a thickness of around 10 μm for DSSCs. Purification of the CNT fabric was performed by heating the samples at high temperature followed by acid treatment.[61] Fibres of DWCNTs with ultra-high purity were obtained from the commercial supplier Dexmat. These fibers are produced by wet-spinning from a liquid crystalline dispersion of high-crystallinity predominantly double-walled carbon nanotubes DWCNTs.

### Device fabrication

After following the standard washing procedure,[62] each electrode was subjected to UV-O$_3$ treatment for 18 min (Model No. 256e220, JelightCompany, Inc). The pre-cleaned FTO substrates were pre-treated by immersion into a 45 mM TiCl$_4$ solution (titanium (IV) chloride solution, 0.09M in 20% HCl from Sigma), in order to suppress the electron recombination, and annealed at 400 °C (30 min). After cooling down, an opaque film of TiO$_2$ paste (18NR-AO Active Opaque Titania Paste) or 2 layers of transparent TiO$_2$ paste (18NR-T Transparent Titania Paste) were doctor bladed on the electrodes. Right after, the electrodes were annealed at 325 °C (5min), 375 °C (5 min), 450 °C (15 min), and at 500 °C (15 min). The resulting films were post-treated with 20 mM TiCl$_4$ aqueous solution for 60 min at 70 °C and re-annealed at 450 °C (60 min), resulting in a thin film of around 10 μm, as measured by a stylus profilometer (KLA-Tencor, Alpha-Step D500). The CNTf-CEs were prepared by press transferring CNTf strips on pre-cleaned FTO glass, condensing with drops of pure ethanol and drying at 70°C for 15 mins, reaching a thickness of 10 μm. For the Pt electrodes used as a reference, a 5 mM solution of hexachloroplatinic acid in isopropanol was drop casted on pre-cleaned FTO substrates and annealed at 400°C for 20min. The sealing of the devices was performed using a spacer of 50 μm thickness. The photoanodes were further sensitized overnight by immersion into a 0.2 mM [(E)-3-(5-(4-(bis (2', 4'-dibutoxy-[1, 1'-biphenyl]-4-yl) amino) phenyl) thiophen-2-yl) -2 cyanoacrylic acid (D35) - Dyenamo] ethanol solution. The cobalt-based electrolyte consisting of 0.25 M [Co(bpy)$_3$][TFSI]$_2$, 0.06 M [Co(bpy)$_3$][TFSI]$_3$, 0.1 M Bis-trifluoromethane-sulfonimide lithium salt (LiTFSI), and 0.5 M 4-tert-butylpyridine (4-TBP) in acetonitrile (ACN), was injected between the electrodes. The symmetrical cell configuration that was used for the electrochemical characterization consisted of two identical electrodes of CNTf or Pt, separated by a spacer

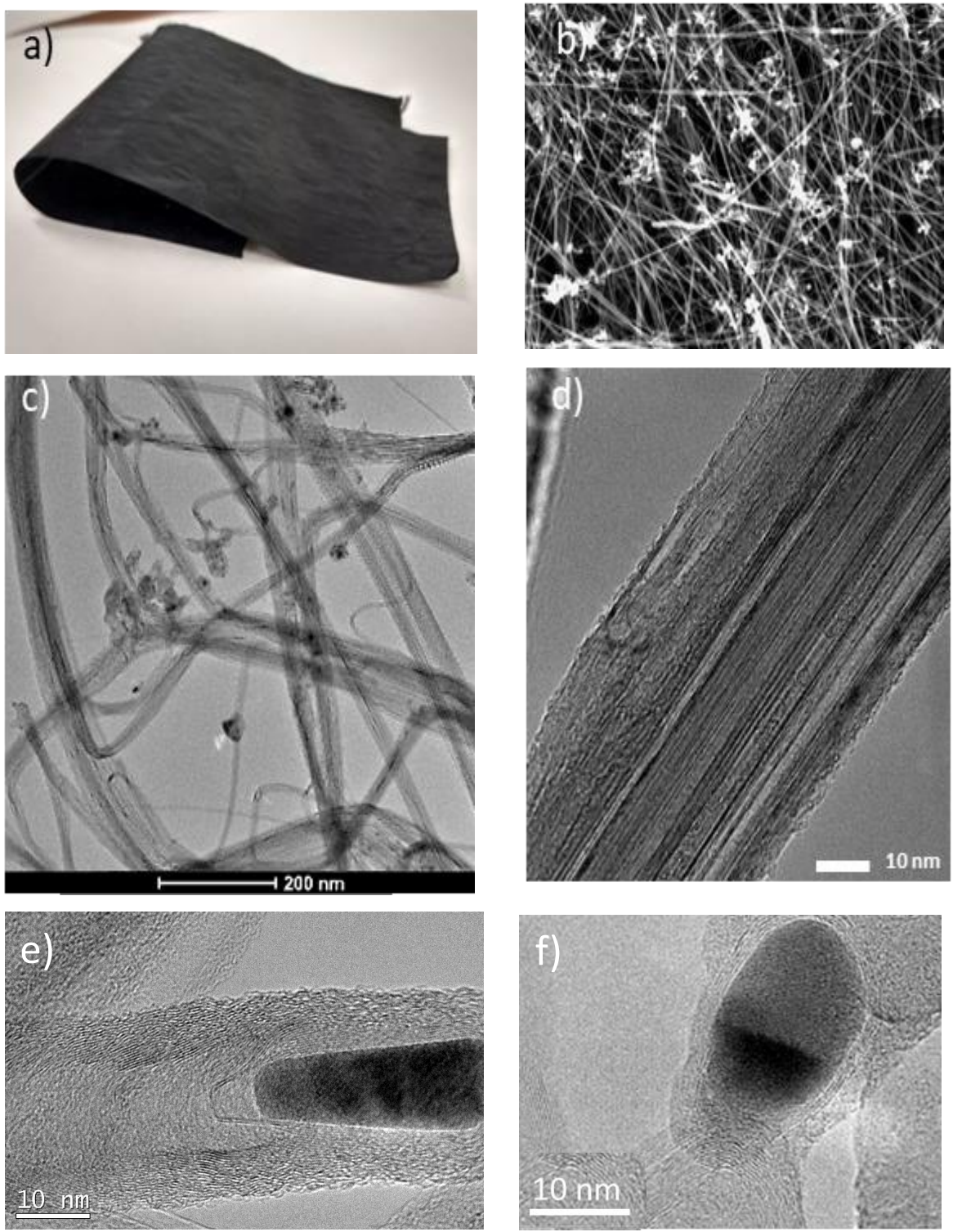

**Figure 1.** Digital photograph (a), SEM (b), and low magnification TEM (c) images showing the large porosity of the CNT and its impurities. High magnification TEM (d) image highlights the high degree of graphitization of the few-layer CNTs. High resolution TEM (HRTEM) of encapsulated Fe impurities (e,f).

(100 µm) and incorporating the cobalt electrolyte. The device figure-of-merit discussed in this work are an average of up to 3 devices, having a standard deviation of *ca*. 0.7%.

**Instrumentation**

Raman analysis was performed on a Renishaw inVia Reflex spectrometer with an $Ar^+$ ion laser ($\lambda$ = 532 nm) excitation source. The laser light was focused on the samples using a 100 objective lens at power of around 1.6 mW in order to avoid sample heating. Spectral deconvolution was carried out by non-linear least square fitting of the Raman peaks to Lorentzian line shapes. Cyclic voltammograms (CV) were recorded on thin layer symmetrical cells made of identical electrodes, filled with the electrolyte, in an electrode/electrolyte/electrode configuration. CV experiments were conducted using an Autolab PGSTAT-30 potentiostat (Ecochemie) with a scan rate of 20 mV/s and a voltage range from -1 to +1 V *versus* electrode. The stability studies were carried out before and after repetitive CV cycles (200) monitoring the changes in the Raman spectra of the CNTfs.

The diffusion coefficient ($D_{Co+3}$) of the reaction determining species ($Co^{+3}$), in the electrolyte bulk has been calculated from the *J-V* curves of the CV assays in the region of the limiting current, using the following equation of Fick's diffusion law:

$$D_{Co^{3+}} = J_{lim}\delta/nFC$$

where *n* = 1 is the number of exchanged electrons during the redox reaction, *C* is the initial concentration of the determining species, *δ* the distance of the electrodes, and *F* the Faraday constant.

The theoretical diffusion constant was calculated using the Stokes-Einstein relation:

$$D_{Co^{3+}} = k_b T/6\pi\rho_{ACN}\, r_{Co^{3+}}$$

with $k_b$ the Boltzman constant, *T* the temperature, $\rho_{ACN}$ the viscosity of *ACN* at 298 K and $r_{Co^{3+}}$ the radius of Co(bpy)$_3{}^{3+}$ [63]. The theoretical current density was calculated with the Randles-Sevcik equation:

$$J = 0.4463nFC\,(nFvD/RT)^{1/2}$$

where n is the number of exchanged electrons, F the Faraday constant, C the concentration, v the scan rate velocity, D the diffusion constant and R the universal gas constant.

Current density-voltage (*J-V*) measurements were performed by illuminating the DSSCs (cell structure: FTO/compact layer/mesoporous TiO$_2$/D35/Co$^{+3}$/Co$^{+2}$ electrolyte/CNTf/FTO) using solar simulated light (1 sun, 1000 W m$^{-2}$) calibrated with a KG5-filtered Silicon reference cell (Newport/Oriel, Model 91150V) from a 150W-Xe source (Oriel), in combination with AM 1.5G optical filter. The active area of the DSSCs was set at 0.09 cm$^2$, using a large black mask in front of the cells in order to avoid any light piping inside the cell. The *J-V* characteristics were recorded using linear sweep voltammetry with the Autolab potentiostat working in a 2-electrode mode at a scan rate of 50 mV/s. Photoelectrochemical and photovoltaic characterization were performed on a batch of at least three cells for each electrode, and the mean value (without a significant deviation) for the obtained results was derived. Results from electrochemical and photoelectrochemical measurements were deduced from the cells that presented performance closest to the average. Electrochemical impedance spectra (EIS) were recorded on the cells using the same potentiostat, equipped with a frequency response analyzer (FRA), at 0 V vs. electrode in dark and light conditions recorded over a frequency range of 100 kHz to 10 mHz. The obtained spectra were fitted with the FRA software provided by the Autolab in terms of appropriate equivalent circuits.

## Results and discussion

### The role of functional groups

The CE material are macrosocpic fabrics of CNTs produced by directly spining a continous fibre from the gas phase during CNT synthesis – see experimental section for more details. **Figure 1** shows that the internal structure of the fabrics consists of an open network of CNT bundles giving rise to mesopores (10-50 nm) and a high specific surface area (with values above 250 m$^2$/g).[64] In addition to CNTs, the samples have non-graphitic carbon impurities and 9 wt.% residual catalyst nanoparticles distributed throughout the fabric and encapuslated along the fibre, as shown in **Figure 1 (e,f)**. These nanoparticles consist of an Fe core (predominanlty metallic γ-Fe) and a continous encapsulating sheel of graphene.[65] They show no preferential faceting. The CNTs are exceptionally long (1 mm), have a few layers, and a relatively high degree of perfection – see **Figure 1d** and Raman spectra in **Figure S1**.

To gain insight into role of surface defects, a set of samples were functionalized through ozonolysis over time.[66,67] The method preserves a highly conducting network, while enabling controlled introduction of oxygen containing groups, such as C-O, O-C=O, COOH and C-OH on the graphitic surface of the material and the concentration of O/C ratio is higher compared to the pristine material.[60,66] Other superficial defects, such as holes and broken layers, appears as a consequence of the transformation of C=C bonds to out of plane C-C bonds in the new functional groups. Ozonolysis also removes surface impurities with no detectable effect on the residual catalyst nanoparticles.[66]n

The degree of functionalization can be readily monitored by Raman spectroscopy: a drop in absolute intensity due to resonance loss, the appearance of the D' mode (double resonance induced by defects and disorder) and the increase of the intensity ratio of the D and G bands ($I_D/I_G$)[68] are the main characteristics. Here, the $I_D/I_G$ increases from 0.64 ± 0.08 for the pristine material (**1**), to 0.80 ± 0.05 (**2**), and to 1.38 ± 0.05 (**3**) for samples functionalised for 2 and 4 minutes, respectively. XPS of functionalized fibres shows an increase in the O1s region, along with an increase In the O1s/C1s ratio, due to the formation of oxygen functional groups. The corresponding changes in surface energy and elemental O/C concentration can be found in the following references[56, 68].

These CNTf samples were transferred onto glass substrates that were applied as electrodes for symmetrical cells with a CNTf/Cobalt-based electrolyte/CNTf configuration – see experimental section for details. The electrocatalytic activity of pristine and functionalized CNTf-CEs was studied using Linear Sweep Voltammetry (LSV) assays. Symmetrical cells with Pt-electrodes are provided for reference purposes. In all cases, the reference electrode was short circuited with the working electrode. Similar to what has been described using iodine-electrolytes,[32-34] LSV curves are almost symmetrical with respect to oxidation/reduction peaks. **Figure 2** displays that **1**-cells present slightly better electrocatalytic behavior towards $Co^{+2} \rightarrow Co^{+3}$ redox reaction than that of Pt-cells as lower redox potentials and similar limiting current densities ($J_{lim}$) are noted – **Table 1**.

This further enhances upon functionalizing the CNTfs. For instance, **2**- and **3**-cells feature redox potential values that hold almost constant, while the $J_{lim}$ and $D_{diff}$ values are almost 2-fold higher than those for **1** and **Pt**-cells – **Table 1**. Interestingly, **2**- and **3**-cells exhibit similar voltammograms, suggesting that the enhanced performance is related to an increase of the surface area due to the formation of new defects and pores.[69] Indeed, the $D_{diff}$ values of *ca.* 4.5 × 10$^{-6}$ cm$^2$/s are very close to the values obtained in literature for highly efficient devices based on liquid electrolytes.[43, 60–65] This suggests that the electrodes are well-optimized; the typical mass transport limitation caused by the bulky nature of the $Co^{+2}/Co^{+3}$ complexes is circumvented with the functionalisation of the CE. Regarding the electrochemical impedance spectrsocopy study (EIS), **1**-cells (Figure S3) presents two semicircles of ≈250Ω and >400Ω in the Nyquist plot centered at 10$^3$-10$^4$ Hz, corresponding to the charge transfer resistance ($R_{ct}$) and at 10$^{-2}$ -1 Hz, corresponding to the bulk diffusion of the electrolkyte ($R_{diff}$), respectively. Those contributions are highly increased after the LSV assys, confirming the dissapeareance of the

**Table 1**. Spectroscopic and electrochemical characterization of the electrodes.

activity. As the functionalization time increases, $R_{ct}$ and $R_{diff}$ remains centered at 10$^4$ Hz and 10$^{-2}$ Hz, with less variation after the LSV assays (**Figure S4**), supporting the stability increase seen in the LSV

In general, CEs must combine high electrocatalytic activity and electrochemical stability; though this has been difficult when catalytic activity originates from defects in carbon-based CEs.[16–18,27,28] In addition, the stability issue is even more critical in cobalt-based electrolytes, as they are prone to degrade.[42,43, 60–65] In this line, the low stability of Pt-cells is manifested by an increase of the redox potential and a decrease of $J_{lim}$ after 10 LSV cycles – **Figures 2** and **3**. This is typically ascribed to the interaction with tert-butyl pyridine (TBP) and the ligand dissociation in ACN[76,77] along with Pt poisoning by air exposure.[42,43, 60–65]

In contrast, **1**-cells feature stable redox potentials over 30 cycles, showing a loss of $J_{lim}$ of <60 % with respect to the fresh electrodes – **Figure 2.** Surprisingly, the stability is further enhanced upon

The enhanced electrochemical stability and the high current density of CNTf-CEs upon functinalization could be realted to two possible mechanisms: i) electrochemical degradation of the electrode promoting the formation of more catalytic centers – *i.e.*, surface defects, and ii) increased number of charge transfer events and weaker interaction of the bipyridine ligand with the functionalized CNTf surface. First, the stability of the CNTf-CEs themselves was studied by Raman spectroscopy – see experimental section for functionalization. For instance, **3**-cells retain stable redox potentials after 100 LSV cycles, while maintaininga high $J_{lim}$.

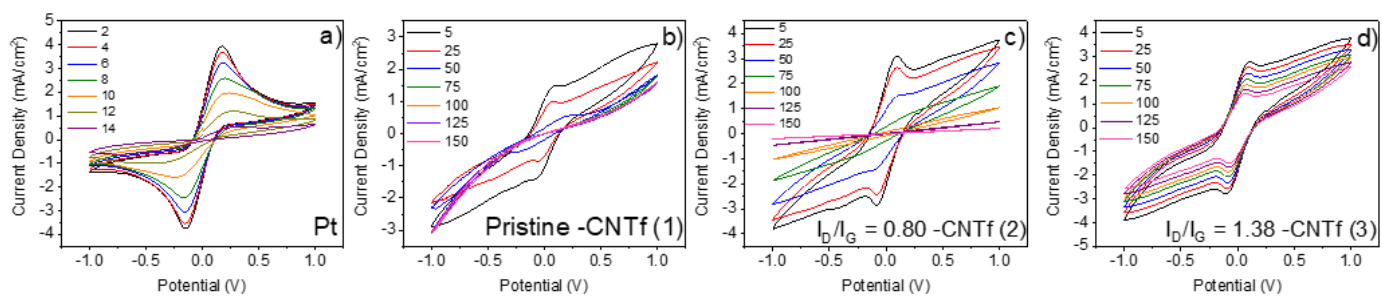

**Figure 2.** LSV assays of symmetric cells with **Pt** (a) and CNTf electrodes: **1** (b), **2** (c) and **3** (d).

**Figure 3** gathers the changes in redox potential and $J_{lim}$ as a details. **Figure S2** shows no significant changes in either peak position or intensity of the Raman spectra features of CNTf-electrodes remain same before and after electrochemical stress, regardless of degree

| Electrodes | $I_D/I_G$* | Impurities content (%) | $V_{red}$ (V) | $V_{ox}$ (V) | $J_{lim}$ (mA/cm$^2$) | $D_{Co^{3+}} \times 10^{-6}$ (cm$^2$/s) |
|---|---|---|---|---|---|---|
| 1 | 0.64 | ≈ 9 | -0.057 | 0.089 | 1.536 | 2.65 |
| 2 | 0.80 | ≈ 9 | -0.079 | 0.101 | 2.577 | 4.45 |
| 3 | 1.38 | ≈ 9 | -0.097 | 0.103 | 2.615 | 4.52 |
| Pt | - | - | -0.145 | 0.179 | 1.517 | 2.62 |

function of the number of cycles ($N_c$), highlighting i) the higher electrochemical stability of CNTf CEs compared to Pt, and ii) the large increase in $J_{lim}$ and its retention upon functionalization. Fitting of the experimental data provides simple metrics to compare the different electrodes – **Table 2**.

The redox potential remains fairly linear and dramatically changes at the onset of the cell degradation. Functionalization reduces the associated exponent, reaching purely planar behavior in the highest functionalization sample – *i.e.,* **3**-cells. $J_{lim}$ is found to decrease linearly with cycle number, but with an increasing slope with reduced functionalization of 0.0102, 0.201, 0.23 mA/cm$^2$/cycle for **3**, **2**, and **1**-cells, respectively.

of functionalization. This evidences the lack of electrode degradation after operation under repetitive cycles, highlighting its exceptional stability. Indeed, the latter is also confirmed, since the redox potential did not change upon functionalization, while **2**- and **3**-cells featured, in addition, similar $J_{lim}$ values – **Table 1**. The second and most likely reason for the increased current density and the electrocatalytic stability is the presence of defects in the CNTF. Along with high surface area, presence of oxygenated functional groups in the ozonized CNTF might attribute to decrease in the activation energy of the electrocatalytic redox process of [Co(bipy)$_3$]$^{3+}$/[Co(bipy)$_3$]$^{2+}$, thereby increasing the current density. Such instances are widely established theoretically as well as experimentally for the iodide/triiodide redox pair.[78,79] On the other hand, the redox reaction of the $[Co(bipy)_3]^{3+}/[Co(bipy)_3]^{2+}$ pair is an outer sphere electron transfer process, where strength of the metal-ligand bond is strongly affected during the electron transfer (Co(II): $(t^2_g)^5(e_g)^2$ to Co(III): $(t^2_g)^6(e_g)^0$). Since this catalytic electron transfer process occurs in the proximity of the graphitic layers of the CNT, it is highly probable that the bipyridine ligands can interact with the CNT surface by π-π interaction and stabilize irreversibly.[80,81] The origin of this process can be compared with the fact of selective separation of small molecules with available π -clouds using CNT by taking advantage of the π electron rich, high surface area of the CNTs.[82,83] Now the possibility of such non-covalent interaction of the bipyridine ligand with the CNT surface decreases when the outermost walls of the CNT contains some defects (e.g. oxygen functional groups) making the graphitic surface very uneven for irreversible stabilization of the bipyridine ligands. This in turn possibly makes the redox process of $[Co(bipy)_3]^{3+}/[Co(bipy)_3]^{2+}$ more reversible on the ozonized CNTF compared to the pristine CNTf. Note that an increase in surface functional groups – i.e. in $I_D/I_G$– produced an exponential increase in $J_{lim}$ (**Figure S5**) and a large increase in stability, but without substantial changes in electrochemical redox potential.

**Table 1.** Empirical dependence of redox potential and limit current density on cycle number.

| Electrode | Oxidation potential $V_{ox} = V_{ox0} - ae^{bN_c}$) | Limit current density ($J_{lim} = J_{lim0} - c \times N_c$) |
|---|---|---|
| 1 | $-0.060 - 2 \times 10^{-5} e^{0.18 \times N_c}$; $R^2$ = 0.98 | $1.59 - 0.023 \times N_c$ ; $R^2$ = 0.95 |
| 2 | $-0.079 - 1 \times 10^{-4} e^{0.10 \times N_c}$; $R^2$ = 0.97 | $2.66 - 0.0201 \times N_c$ ; $R^2$ = 0.97 |
| 3 | $-0.096 - 2.2 \times 10^{-4} \times N_c$; $R^2$ = 0.80 | $2.66 - 0.0102 \times N_c$ ; $R^2$ = 0.99 |

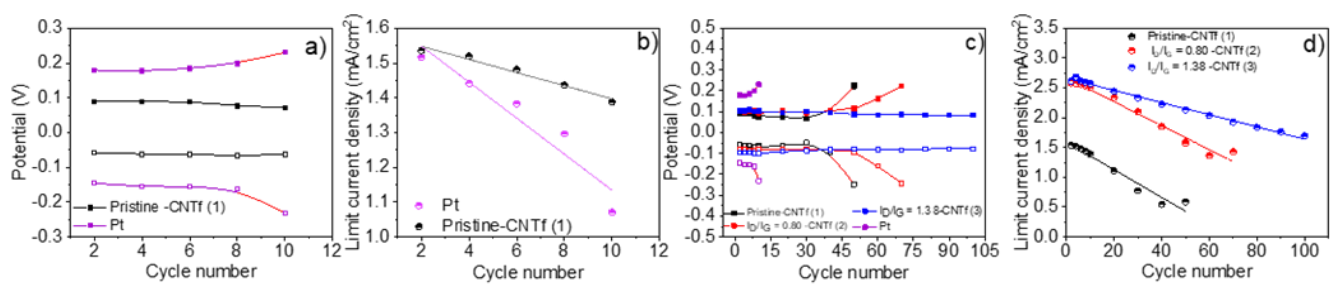

**Figure 3.** Reduction (empty squares) and oxidation (filled squares) potentials (a and c) and limit current densities (b and d) versus number of cycles of symmetric cells with **Pt** (purple), **1** (black), **2** (red) and **3** (blue). Lines in b) and d) correspond to the fittings of the limit current densities.

**Metal impurities effect**

We turned to investigate the impact of the metal impurities on the electrocatalytic behavior. The amount of impurities was reduced from 9% wt. to 3 % wt. through acid treatment, followed by electrochemical etching – **Figure S6**.[61] This procedure was adjusted to also produce a similar functionalization degree to that of the above best CNTf electrode (**3**) – **Figure S7**. This sample is labelled **3-3%**. Further comparison was also made with a commercial CNTf sample with effectively 0% wt. of metal impurities, which was also oxidized to reach the same functionalization degree – **Figure S8**. This was labelled **3-0%**. Through LSV and EIS tests on symmetric cells with these samples, we could, therefore, study the effect of the residual metallic catalyst content in the range 9, 3, and 0% wt. at nearly the same degree of functionalization.

As shown in **Figure 4**, the decrease of the amount of impurities significantly impact the electrocatalytic behavior of the CNTf-electrodes. In short, **3-3%**-cells showed a 2-fold higher redox potential compared to **3**-cells, indicating that the metal impurities presented along the fibre plays a key role as catalytic center of the redox reaction. Importantly, the redox potential is stable over 100 LSV cycles while both, initial and loss $J_{lim}$ values, are similar to those of **3**. This points out that the functionalization is key towards enhanced electrocatalytic stabilities and current densities, while the Fe impurities are responsible of the electrocatylic activity. This is further established by the featureless LSV curves of **3-0%**-cells – **Figure 4**, in which the Fe impurities are not present in the CNTfs. Importantly, this CNTf elecrodes did not show any electrocatalytic activity in both pristine and functionalized forms – **Figure S9**. Regarding EIS tests (**Figure S10**), **3-3%** shows similar resistance values to those obtained with the functionalized samples 2 and 3, while the functionalized 3-0% symmetric cell presents $10^6 \Omega$, showing the no-activity of this cell.

This confirms that the type and amount of metallic impurities is key towards the future optimization of CNTf electrodes for energy-related application.

**Solar cells characterization and performance**

Having determined the best configuration for CNTf-elecrodes, we compare the performance of DSSCs with two different CEs: Pt as reference and the optimum CNTf sample (**3**) – *i.e.*, highly functionalized and 9 %wt. metallic impurities. The DSSCs were fabricated with $TiO_2$ mesoporous electrodes sensitized with an organic dye (D35) as photoanodes and liquid cobalt-based electrolyte – see experimental section for details.[70–72] **Figures 5**

Devices using **Pt**-CEs exhibit short circuit current ($J_{sc}$) values of 1.55 mA/cm$^2$ and $V_{OC}$ values of 0.85 V. The $V_{OC}$ value is close to the theoretical one predicted for this cobalt-based electrolyte,[42] while the IPCE matches the absorption of the dye, indicating that photogenerated currents are related to the photoexcitation of the dye. This goes along with fill factor (*FF*) values above 0.7, highlighting the quality of the solar cells.

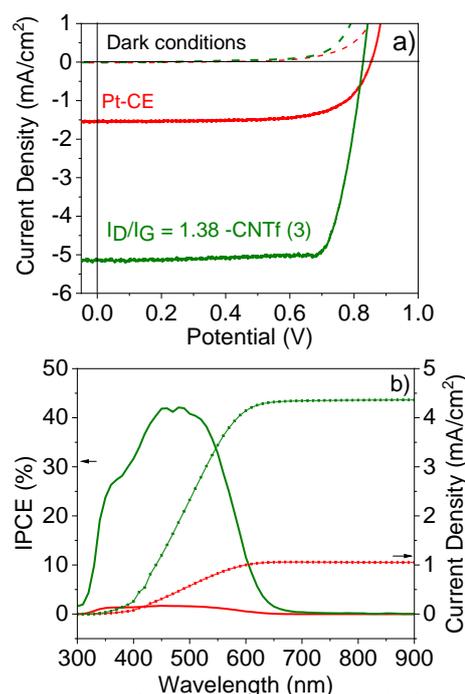

**Figure 5.** J-V characteristics (top) of the **3**- (green) and **Pt**- (red) DSSCs under 1-sun illumination (solid line) and dark conditions (dashed lines) as well as IPCE and integrated current density (bottom).

The devices with **3**-CEs feature similar device quality (*FF* = 70*)*, but they exhibit a 5-fold increase in $J_{sc}$ values. This is clearly in agreement with results on symmtric cells. In line with devices using iodine-basd electrolytes,[32-34] the $V_{oc}$ is, however, reduced. This could be realted to changes in the redox potential of the cobalt-based electrolytes due to different concentration gradients between the bulk electrolye and the electrode surface as well as changes in the pH.[84] Overall, the device efficiency increases from 1.5 % (**Pt**-DSSCs) to 2 % (**3**-DSSCs), highlighting the superior performance of the CNTf-displays the J-V curves and incident photon-to-current conversion efficiencies (IPCE) obtained under 1 sun AM 1.5G and dark conditions.

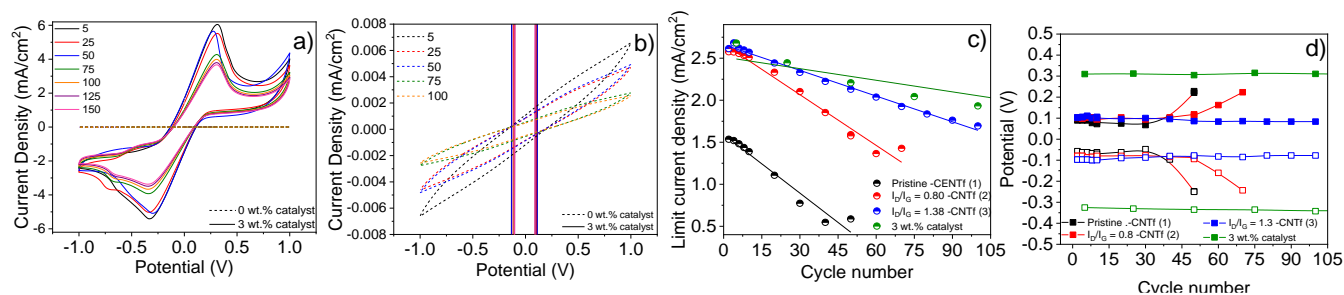

**Figure 4.** LSV of symmetric cells with **3-3%** (solid line) and **3-0%** (dashed line) and an inset to better show the current density scales (b). Limit current densities (c) and reduction (empty squares) and oxidation (filled square) potentials (d) versus Cycle number of symmetric cells with **1** (black), **2** (red), **3** (blue), and **3-3%** (green) are shown. Lines in c) are fittings of the limit current densities.

CE for cobalt-based electrolytes.

EIS measurements were also carried out to gain further insight into the different processes taking place in the solar cells – **Figure**

**S11**. The Nyquist plots consisted of only one or two semicircles centered at around 1-100 Hz and 1000-10000 Hz. The former is related to charge transport resistance of the photoanode, while the latter relates to the Warburg ion resistance. Therefore, we were unable to determine the resistance associated to the charge transfer (Rct) process at the electrolyte/CNTf-CE interface. This problem has already been encountered in devices based on iodine-based electrolytes.[36]

In line with the LSV assays, **Pt**-DSSCs showed a loss in $J_{sc}$ of >90% after 1 week, while $V_{oc}$ and $FF$ hold constant after 3 weeks – **Figure 6**. In stark contrast, **3**-DSSCs exhibited a notably superior stability. After 3 weeks under working conditions, both, $V_{oc}$ and $FF$, remained constant, while $J_{sc}$ slightly reduced by <10%.

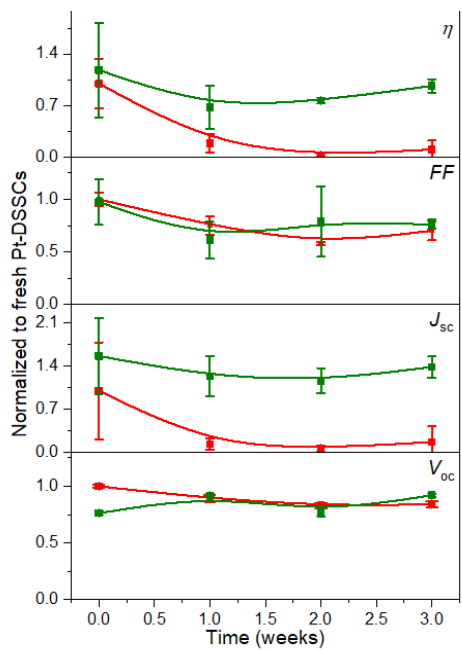

**Figure 6.** Changes of the figures-of-merit-of **3**- (green) and **Pt**- (red) DSSCs over time under operation.

## Conclusions

We have deciphered the origin of electrocatalytic behavior of CNTf electrodes with the aim to use them as alternative to Pt-CEs in DSSCs with a $Co^{+2}/Co^{+3}$ redox couple electrolyte. Similar to what has been observed in iodide-based electrolytes, the CNTf-CEs feature a significant catalytic activity towards $Co^{+2}/Co^{+3}$ redox couple electrolytes that can be optimized in terms of its functionalization and metal impurities content. On one hand, upon CNTf functionalization through ozonolysis ($I_D/I_G$=1.38), higher limit current densities and stabilities are realized in symetric cells. On the other hand, metallic Fe impurities are key towards the catalytic behavior that is inexistant for CNTf without metal catalyst and superior for CNTf electrodes with the higher amount of 9 %wt. These findings were used to optimize the CNTf-electrodes for DSSCs. Here, devices prepared with the optimized CNTf-CEs clearly ouperformed those with Pt-CEs, showing, in addition, operation stabilities of several weeks.

Building on recent studies focused on devices with standard $I^-/I_3^-$ redox couple,[31-38] we provide new insights on the most important aspects towards optimizing CNTf-electrodes for solar energy conversion devices using other redox electrolytes. Thus, we strongly believe that this work paves the way for using other electrolytes in CNTf-based DSSCs, in general, and other solar-driven energy applications using CNTf-electrodes, in particular.

## Conflicts of interest

There are no conflicts to declare.

## Acknowledgements

Financial support is acknowledged from the European Union Seventh Framework Program under grant agreement 678565 (ERC-STEM), the Clean Sky Joint Undertaking 2, Horizon 2020 under Grant Agreement Number 738085 (SORCERER), and from MINECO (MAT2015-62584 ERC, RyC-2014-15115, Spain) and CAM MAD2D project (S2013/MIT-3007). A. M. M. and R.D.C. further of the Consejería de Educación, Juventud y Deporte – Comunidad de Madrid with reference number 2016-T1/IND-1463". R. D. C. also acknowledges the Spanish MINECO for the Ramon y Cajal program (RYC-2016-20891), the Europa Excelencia program (ERC2019-092825), and HYNANOSC (RTI2018-099504-A-C22),as well as the 2018 Leonardo Grant for Researchers and Cultural Creators from BBVA Foundation, the FOTOART-CM project funded by Madrid region under program P2018/NMT-4367.

## Notes and references

1. G. Centi and S. Perathoner, *ChemSusChem*, 2011, **4**, 913–925.
2. P.-X. Hou, J. Du, C. Liu, W. Ren, E. I. Kauppinen and H.-M. Cheng, *MRS Bull.*, 2017, **42**, 825–833.
3. M. Melchionna, S. Marchesan, M. Prato and P. Fornasiero, *Catal. Sci. Technol.*, 2015, **5**, 3859–3875.
4. J. H. Hafner, M. J. Bronikowski, B. R. Azamian, P. Nikolaev, A. G. Rinzler, D. T. Colbert, K. A. Smith and R. E. Smalley, *Chemical Physics Letters*, 1998, **296**, 195–202.
5. D. Takagi, Y. Homma, H. Hibino, S. Suzuki and Y. Kobayashi, *Nano Lett.*, 2006, **6**, 2642–2645.
6. S. Ahmad, Y. Liao, A. Hussain, Q. Zhang, E.-X. Ding, H. Jiang and E. I. Kauppinen, *Carbon*, 2019, **149**, 318–327.
7. M. Pumera, A. Ambrosi and E. L. K. Chng, *Chem. Sci.*, 2012, **3**, 3347.
8. C. E. Banks, A. Crossley, C. Salter, S. J. Wilkins and R. G. Compton, *Angew. Chem. Int. Ed.*, 2006, **45**, 2533–2537.
9. M. Pumera and Y. Miyahara, *Nanoscale*, 2009, **1**, 260.
10. K. Aitola, J. Halme, N. Halonen, A. Kaskela, M. Toivola, A. G. Nasibulin, K. Kordás, G. Tóth, E. I. Kauppinen and P. D. Lund, *Thin Solid Films*, 2011, **519**, 8125–8134.
11. Y. Wang, M. Yi, K. Wang and S. Song, *Chinese Journal of Catalysis*, 2019, **40**, 523–533.

12 P. Dong, C. L. Pint, M. Hainey, F. Mirri, Y. Zhan, J. Zhang, M. Pasquali, R. H. Hauge, R. Verduzco, M. Jiang, H. Lin and J. Lou, *ACS Appl. Mater. Interfaces*, 2011, **3**, 3157–3161.
13 S. Huang, Z. Yang, L. Zhang, R. He, T. Chen, Z. Cai, Y. Luo, H. Lin, H. Cao, X. Zhu and H. Peng, *J. Mater. Chem.*, 2012, **22**, 16833.
14 K. Kakiage, Y. Aoyama, T. Yano, K. Oya, J. Fujisawa and M. Hanaya, *Chemical Communications*, 2015, **51**, 15894–15897.
15 M. Freitag, J. Teuscher, Y. Saygili, X. Zhang, F. Giordano, P. Liska, J. Hua, S. M. Zakeeruddin, J.-E. Moser, M. Grätzel and A. Hagfeldt, *Nature Photon*, 2017, **11**, 372–378.
16 S. Yun, P. D. Lund and A. Hinsch, *Energy Environ. Sci.*, 2015, **8**, 3495–3514.
17 F. Lodermeyer, R. D. Costa and D. M. Guldi, *Adv. Energy Mater.*, 2017, **7**, 1601883.
18 R. D. Costa, F. Lodermeyer, R. Casillas and D. M. Guldi, *Energy Environ. Sci.*, 2014, **7**, 1281.
19 H. C. Weerasinghe, F. Huang and Y.-B. Cheng, *Nano Energy*, 2013, **2**, 174–189.
20 C. Wu, B. Chen, X. Zheng and S. Priya, *Solar Energy Materials and Solar Cells*, 2016, **157**, 438–446.
21 G. Galimberti, S. Pagliara, S. Ponzoni, S. Dal Conte, F. Cilento, G. Ferrini, S. Hofmann, M. Arshad, C. Cepek and F. Parmigiani, *Carbon*, 2011, **49**, 5246–5252.
22 G. Galimberti, S. Ponzoni, A. Cartella, M. T. Cole, S. Hofmann, C. Cepek, G. Ferrini and S. Pagliara, *Carbon*, 2013, **57**, 50–58.
23 M. Grechko, Y. Ye, R. D. Mehlenbacher, T. J. McDonough, M.-Y. Wu, R. M. Jacobberger, M. S. Arnold and M. T. Zanni, *ACS Nano*, 2014, **8**, 5383–5394.
24 R. D. Mehlenbacher, T. J. McDonough, M. Grechko, M.-Y. Wu, M. S. Arnold and M. T. Zanni, *Nature Communications*, 2015, **6**, 6732.
25 J.-H. Han, G. L. C. Paulus, R. Maruyama, D. A. Heller, W.-J. Kim, P. W. Barone, C. Y. Lee, J. H. Choi, M.-H. Ham, C. Song, C. Fantini and M. S. Strano, *Nature Materials*, 2010, **9**, 833.
26 T. Dürkop, S. A. Getty, E. Cobas and M. S. Fuhrer, *Nano Letters*, 2004, **4**, 35–39.
27 A. Jorio, G. Dresselhaus and M. S. Dresselhaus, Eds., *Carbon nanotubes: advanced topics in the synthesis, structure, properties, and applications*, Springer, Berlin ; New York, 2008.
28 M. Chen and L.-L. Shao, *Chemical Engineering Journal*, 2016, **304**, 629–645.
29 M. Z. Iqbal and S. Khan, *Solar Energy*, 2018, **160**, 130–152.
30 J.-P. Salvetat-Delmotte and A. Rubio, *Carbon*, 2002, **40**, 1729–1734.
31 X. Fu, H. Sun, S. Xie, J. Zhang, Z. Pan, M. Liao, L. Xu, Z. Li, B. Wang, X. Sun and H. Peng, *J. Mater. Chem. A*, 2018, **6**, 45–51.
32 W. J. Lee, E. Ramasamy, D. Y. Lee and J. S. Song, *ACS Appl. Mater. Interfaces*, 2009, **1**, 1145–1149.
33 J. G. Nam, Y. J. Park, B. S. Kim and J. S. Lee, *Scripta Materialia*, 2010, **62**, 148–150.
34 Z. Yang, T. Chen, R. He, G. Guan, H. Li, L. Qiu and H. Peng, *Adv. Mater.*, 2011, **23**, 5436–5439.
35 T. Chen, L. Qiu, Z. Cai, F. Gong, Z. Yang, Z. Wang and H. Peng, *Nano Lett.*, 2012, **12**, 2568–2572.
36 A. Monreal-Bernal, J. J. Vilatela and R. D. Costa, *Carbon*, 2019, **141**, 488–496.
37 P. Poudel, L. Zhang, P. Joshi, S. Venkatesan, H. Fong and Q. Qiao, *Nanoscale*, 2012, **4**, 4726.
38 A. Aboagye, H. Elbohy, A. D. Kelkar, Q. Qiao, J. Zai, X. Qian and L. Zhang, *Nano Energy*, 2015, **11**, 550–556.
39 H. Elbohy, A. Aboagye, S. Sigdel, Q. Wang, M. H. Sayyad, L. Zhang and Q. Qiao, *J. Mater. Chem. A*, 2015, **3**, 17721–17727.
40 P. Joshi, Z. Zhou, P. Poudel, A. Thapa, X.-F. Wu and Q. Qiao, *Nanoscale*, 2012, **4**, 5659.
41 Y. Xiao, J. Wu, J. Lin, G. Yue, J. Lin, M. Huang, Y. Huang, Z. Lan and L. Fan, *J. Mater. Chem. A*, 2013, **1**, 13885.
42 S.-Y. Tai, C.-J. Liu, S.-W. Chou, F. S.-S. Chien, J.-Y. Lin and T.-W. Lin, *J. Mater. Chem.*, 2012, **22**, 24753.
43 G. Yue, J. Wu, J.-Y. Lin, Y. Xiao, S.-Y. Tai, J. Lin, M. Huang and Z. Lan, *Carbon*, 2013, **55**, 1–9.
44 Z. Zhou, S. Sigdel, J. Gong, B. Vaagensmith, H. Elbohy, H. Yang, S. Krishnan, X.-F. Wu and Q. Qiao, *Nano Energy*, 2016, **22**, 558–563.
45 X. Fu, H. Sun, S. Xie, J. Zhang, Z. Pan, M. Liao, L. Xu, Z. Li, B. Wang, X. Sun and H. Peng, *J. Mater. Chem. A*, 2018, **6**, 45–51.
46 Z. Yang, J. Deng, X. Sun, H. Li and H. Peng, *Adv. Mater.*, 2014, **26**, 2643–2647.
47 S. Pan, H. Lin, J. Deng, P. Chen, X. Chen, Z. Yang and H. Peng, *Adv. Energy Mater.*, 2015, **5**, 1401438.
48 H. Sun, X. You, J. Deng, X. Chen, Z. Yang, P. Chen, X. Fang and H. Peng, *Angew. Chem. Int. Ed.*, 2014, **53**, 6664–6668.
49 S. N. Habisreutinger, R. J. Nicholas and H. J. Snaith, *Adv. Energy Mater.*, 2017, **7**, 1601839.
50 S. Jiao, J. Du, Z. Du, D. Long, W. Jiang, Z. Pan, Y. Li and X. Zhong, *J. Phys. Chem. Lett.*, 2017, **8**, 559–564.
51 A. Yella, S. Mathew, S. Aghazada, P. Comte, M. Grätzel and M. K. Nazeeruddin, *J. Mater. Chem. C*, 2017, **5**, 2833–2843.
52 F. Bella, S. Galliano, C. Gerbaldi and G. Viscardi, *Energies*, 2016, **9**, 384.
53 L. Kavan, Y. Saygili, M. Freitag, S. M. Zakeeruddin, A. Hagfeldt and M. Grätzel, *Electrochimica Acta*, 2017, **227**, 194–202.
54 H. Wu, Z. Lv, Z. Chu, D. Wang, S. Hou and D. Zou, *J. Mater. Chem.*, 2011, **21**, 14815–14820.
55 K. Meng, P. K. Surolia, O. Byrne and K. R. Thampi, *Journal of Power Sources*, 2015, **275**, 681–687.
56 G. Liu, X. Li, H. Wang, Y. Rong, Z. Ku, M. Xu, L. Liu, M. Hu, Y. Yang and H. Han, *Carbon*, 2013, **53**, 11–18.
57 C. V. V. M. Gopi, S. Singh, A. Eswar Reddy and H.-J. Kim, *ACS Appl. Mater. Interfaces*, 2018, **10**, 10036–10042.
58 L. Wang, J. Feng, Y. Tong and J. Liang, *International Journal of Hydrogen Energy*, 2019, **44**, 128–135.
59 Y.-L. Li, *Science*, 2004, **304**, 276–278.
60 B. Alemán, M. Vila and J. J. Vilatela, *Phys. Status Solidi A*, 2018, **215**, 1800187.
61 V.-D. Dao and H.-S. Choi, *Data in Brief*, 2018, **20**, 1153–1159.
62 S. Ito, T. N. Murakami, P. Comte, P. Liska, C. Grätzel, M. K. Nazeeruddin and M. Grätzel, *Thin Solid Films*, 2008, **516**, 4613–4619.
63 N. Yaghoobi Nia, P. Farahani, H. Sabzyan, M. Zendehdel and M. Oftadeh, *Physical Chemistry Chemical Physics*, 2014, **16**, 11481.
64 E. Senokos, V. Reguero, J. Palma, J. J. Vilatela and R. Marcilla, *Nanoscale*, 2016, **8**, 3620–3628.
65 B. Alemán, R. Ranchal, V. Reguero, B. Mas and J. J. Vilatela, *J. Mater. Chem. C*, 2017, **5**, 5544–5550.
66 D. Iglesias, E. Senokos, B. Alemán, L. Cabana, C. Navío, R. Marcilla, M. Prato, J. J. Vilatela and S. Marchesan, *ACS Applied Materials & Interfaces*, 2018, **10**, 5760–5770.

67  J. M. Simmons, B. M. Nichols, S. E. Baker, M. S. Marcus, O. M. Castellini, C.-S. Lee, R. J. Hamers and M. A. Eriksson, *J. Phys. Chem. B*, 2006, **110**, 7113–7118.
68  E. Senokos, M. Rana, C. Santos, R. Marcilla and J. J. Vilatela, *Carbon*, 2019, **142**, 599–609.
69  H. Yue, V. Reguero, E. Senokos, A. Monreal-Bernal, B. Mas, J. P. Fernández-Blázquez, R. Marcilla and J. J. Vilatela, *Carbon*, 2017, **122**, 47–53.
70  D. Kumar and K.-T. Wong, *Materials Today Energy*, 2017, **5**, 243–279.
71  S. M. Feldt, E. A. Gibson, E. Gabrielsson, L. Sun, G. Boschloo and A. Hagfeldt, *J. Am. Chem. Soc.*, 2010, **132**, 16714–16724.
72  X. Jiang, K. M. Karlsson, E. Gabrielsson, E. M. J. Johansson, M. Quintana, M. Karlsson, L. Sun, G. Boschloo and A. Hagfeldt, *Adv. Funct. Mater.*, 2011, **21**, 2944–2952.
73  Y. Hao, Y. Saygili, J. Cong, A. Eriksson, W. Yang, J. Zhang, E. Polanski, K. Nonomura, S. M. Zakeeruddin, M. Grätzel, A. Hagfeldt and G. Boschloo, *ACS Appl. Mater. Interfaces*, 2016, **8**, 32797–32804.
74  W. Yang, Y. Hao, P. Ghamgosar and G. Boschloo, *Electrochimica Acta*, 2016, **213**, 879–886.
75  X. L. Zhang, W. Huang, A. Gu, W. Xiang, F. Huang, Z. X. Guo, Y.-B. Cheng and L. Spiccia, *J. Mater. Chem. C*, 2017, **5**, 4875–4883.
76  S. M. Feldt, P. W. Lohse, F. Kessler, M. K. Nazeeruddin, M. Grätzel, G. Boschloo and A. Hagfeldt, *Phys. Chem. Chem. Phys.*, 2013, **15**, 7087.
77  J. T. Kirner and C. M. Elliott, *J. Phys. Chem. C*, 2015, **119**, 17502–17514.
78  J. Hong, C. Yu, X. Song, X. Meng, H. Huang, C. Zhao, X. Han, Z. Wang and J. Qiu, *ACS Sustainable Chem. Eng.*, 2019, **7**, 7527–7534.
79  J. D. Roy-Mayhew, D. J. Bozym, C. Punckt and I. A. Aksay, *ACS Nano*, 2010, **4**, 6203–6211.
80  G. Tuci, A. Rossin, L. Luconi, C. Pham-Huu, S. Cicchi, H. Ba and G. Giambastiani, *Catal. Sci. Technol.*, 2017, **7**, 5833–5837.
81  E. Cohen, H. Dodiuk, A. Ophir, S. Kenig, C. Barry and J. Mead, *Composites Science and Technology*, 2013, **79**, 133–139.
82  B. Zhao, H. Liang, D. Han, D. Qiu and S. Chen, *Separation Science and Technology*, 2007, **42**, 3419–3427.
83  R. Q. Long and R. T. Yang, *J. Am. Chem. Soc.*, 2001, **123**, 2058–2059.
84  T. N. Murakami, S. Ito, Q. Wang, Md. K. Nazeeruddin, T. Bessho, I. Cesar, P. Liska, R. Humphry-Baker, P. Comte, P. Péchy and M. Grätzel, *J. Electrochem. Soc.*, 2006, **153**, A2255.